\documentclass[twocolumn]{jpsj3}

\usepackage{color}
\usepackage{times}
\usepackage{graphicx}

\title{Anomalous Isotope Effect in Rattling-Induced Superconductor}

\author{Kunihiro {\sc Oshiba}$^{1}$ and Takashi {\sc Hotta}$^{1,2}$}

\inst{$^{1}$Department of Physics, Tokyo Metropolitan University,
Hachioji, Tokyo 192-0397, Japan \\
$^{2}$Advanced Science Research Center, Japan Atomic Energy Agency,
Tokai, Ibaraki 319-1195, Japan}

\recdate{\today}

\abst{
In order to clarify that the Cooper pair in $\beta$-pyrochlore oxides
is mediated by anharmonic oscillation of guest atom, i.e., {\it rattling},
we propose an experiment to detect anomalous isotope effect.
In the formula of $T_{\rm c} \propto M^{-\eta}$,
where $T_{\rm c}$ is superconducting transition temperature
and $M$ denotes mass of the oscillator,
it is found that the exponent $\eta$ is increased with the increase of
anharmonicity of a potential for the guest atom.
We predict that $\eta$ becomes larger than $1/2$ in rattling-induced superconductor,
in sharp contrast to $\eta=1/2$ for weak-coupling superconductivity
due to harmonic phonons and $\eta<1/2$ for strong-coupling superconductivity
with the inclusion of the effect of Coulomb interaction.
}

\kword{Isotope effect, rattling, superconductivity, $\beta$-pyrochlore}

\begin{document}
\maketitle

\section{Introduction}

Recently, exotic magnetism and novel superconductivity have attracted
much attention in the research field of condensed matter physics.
\cite{JPSJ1,JPSJ2,JPSJ3,Review0,Review1,Review2,Review3,skut}
In particular, strongly correlated electron systems with cage structure
have been focused both from experimental and theoretical viewpoints.
In such cage-structure materials, a guest atom contained in the cage
feels a highly anharmonic potential and it oscillates
with relatively large amplitude in comparison with
that of the lattice vibration in metals.
Such oscillation of the atom in the cage is called {\it rattling},
which is considered to be an origin of interesting physical properties
of cage-structure compounds.

Among several interesting phenomena in cage-structure materials,
since the discovery of superconductivity with relatively
high superconducting transition temperature $T_{\rm c}$
in $\beta$-pyrochlore oxides AOs$_{2}$O$_{6}$ (A = K, Rb, and Cs),
\cite{Yonezawa1,Yonezawa2,Yonezawa3,Hiroi1,Kazakov,Bruhwiler,Bruhwiler2}
phonon-mediated superconductivity has attracted renewed attention
from the viewpoint of anharmonicity.
It has been observed that $T_{\rm c}$ increases with the decrease of radius
of A ion:
$T_{\rm c}$=9.6K for A=K, $T_{\rm c}$=6.4K for A=Rb, and $T_{\rm c}$=3.25K
for A=Cs.\cite{Hiroi2}
The difference in $T_{\rm c}$ has been considered to originate from
the anharmonic oscillation of A ion.
In fact, the anharmonicity of the potential for A ion has been found
to be enhanced, when we change A ion in the order of Cs, Rb, and K
due to the first-principles calculations.\cite{Kunes}

As for the mechanism of superconductivity in $\beta$-pyrochlore oxides,
theoretical investigations have been done.
Hattori and Tsunetsugu have investigated a realistic model including
three dimensional anharmonic phonons with tetrahedral symmetry and
have confirmed that $T_{\rm c}$ is strongly enhanced
with increasing the third-order
anharmonicity of the potential.\cite{Hattori1,Hattori2}
Chang {\it et al.} have discussed the superconductivity
by using the strong-coupling approach in the anharmonic phonon model
including fourth-order terms.\cite{Chang}
The present authors have revealed the anharmonicity-controlled
strong-coupling tendency for superconductivity induced by rattling
from the analysis of the anharmonic Holstein model
in the strong-coupling theory.\cite{Oshiba-Proc,Oshiba}

From these experimental and theoretical efforts,
it has been gradually recognized that the superconductivity in
$\beta$-pyrochlore oxides is induced
by anharmonic oscillation of alkali atom contained in a cage
composed of oxygen and osmium.
However, it is unsatisfactorily clarified how the superconductivity
induced by rattling is different from the conventional superconductivity
due to harmonic phonons.
In particular, it is considered to be important to confirm
the evidence for the Cooper-pair formation due to rattling.
Concerning this issue, we hit upon an idea to exploit isotope effect.

Now we recall the famous formula
$T_{\rm c}=1.13 \omega e^{-1/\lambda}$
in the Bardeen-Cooper-Schrieffer theory,\cite{BCS}
where $\omega$ denotes a characteristic phonon energy and $\lambda$ indicates
a non-dimensional electron-phonon coupling constant.
For harmonic phonons, in general, $\lambda$ does not depend on the mass of
oscillator $M$ and then,
it is enough to consider the $M$ dependence of $\omega$.
Since $\omega$ is in proportion to $M^{-1/2}$, we express the relation
between $T_{\rm c}$ and $M$ as $T_{\rm c} \propto M^{-\eta}$ with $\eta=1/2$
for conventional superconductors mediated by harmonic phonons.
In actual experiments on Hg,\cite{Hg1, Hg2} it has been clearly shown that
$T_{\rm c}$ is in proportion to $M^{-1/2}$,
leading to the evidence of phonon-mediated Cooper pair.
However, if we cannot ignore the $M$ dependence of electron attraction
mediated by anharmonic phonons,
there should occur significant deviation of $\eta$ from $1/2$,
which can provide an evidence of rattling-induced superconductivity.

Note that the effect of anharmonic oscillation on $\eta$ was previously
discussed in the research of high-$T_{\rm c}$ cuprate superconductors.
For instance, a model for anharmonic oscillation of oxygen was investigated,
\cite{YBCO,Drechsler,Crespi1} for a possible explanation of
the small exponent $\eta$ of high-$T_{\rm c}$ cuprates.
For La$_{\rm 2-x}$Sr$_{\rm x}$CuO$_4$, there was a trial
to understand the anomalous value of $\eta$
which was larger or smaller than $1/2$ due to the inclusion of
the anharmonic potential.\cite{LaSrCuO, Crespi2}
However, for high-$T_{\rm c}$ cuprates, electron correlation is considered to
play the primary role for the emergence of anisotropic superconductivity.
The research of anomalous isotope effect in cuprates is essentially
different from the purpose of the present paper
to clarify the evidence of Cooper-pair formation due to anharmonic phonons.

In this paper, we evaluate the exponent $\eta$ by applying
the strong-coupling Migdal-Eliashberg theory~\cite{Migdal,Eliashberg}
for rattling-induced superconductor. 
From the analyses of the anharmonic Holstein model,
we find anomalous isotope effect with $\eta > 1/2$,
in sharp contrast to the decrease of $\eta$ from $1/2$
due to the effect of Coulomb interaction
in strong-coupling superconductivity.
It is confirmed that the origin of anomalous isotope effect is certainly
the anharmonicity of potential,
leading to the conclusion that $\eta > 1/2$ can be
the evidence of superconductivity induced by anharmonic phonons.
We propose an experiment on the isotope effect in order to clarify
a key role of rattling in $\beta$-pyrochlore oxides.

The organization of this paper is as follows.
In Sec.~2, we show the anharmonic Holstein Hamiltonian
and explain the model potential for anharmonic oscillation.
We also provide the brief explanation of the formulation of
the Migdal-Eliashberg theory to evaluate
superconducting transition temperature $T_{\rm c}$.
In Sec.~3, we exhibit our calculated results on $T_{\rm c}$
and the values of $\eta$.
We also discuss $T_{\rm c}$ and $\eta$ on the basis of
the McMillan formula.
It is emphasized that the anomalous value of $\eta$
larger than $1/2$ certainly originates from the anharmonicity.
Finally, in Sec.~4, we briefly discuss the effect of the Coulomb interaction on $\eta$ and summarize this paper.
Throughout this paper, we use such units as $\hbar=k_{\rm B}=1$.

\section{Model and Formulation}

\subsection{Anharmonic Holstein model}

In this paper, we consider the Holstein model in which conduction electrons
are coupled with anharmonic local oscillations.
The model is given by
\begin{equation}
 \label{Hamiltonian}
  H = \sum_{\mib{k},\sigma} \varepsilon_{\mib{k}}
      c^{\dagger}_{\mib{k}\sigma} c_{\mib{k}\sigma}
    + \sum_{\mib{i}}\left[ H^{(1)}_{\mib{i}} + H^{(2)}_{\mib{i}} \right],
\end{equation}
where $\mib{k}$ is momentum of electron, $\varepsilon_{\mib{k}}$ denotes
the energy of conduction electron, $\sigma$ is an electron spin, 
$c_{\mib{k}\sigma}$ is an annihilation operator of electron with $\mib{k}$
and $\sigma$, and $\mib{i}$ denotes atomic site.
Throughout this paper, we consider half-filling case and the electron
bandwidth $W$ is set as unity for an energy unit.

In eq.~(\ref{Hamiltonian}), $H^{(1)}_{\mib{i}}$ and $H^{(2)}_{\mib{i}}$,
respectively, denote electron-vibration coupling and vibration terms
at site $\mib{i}$, expressed by
\begin{equation}
  \label{h1}
  H^{(1)}_{\mib{i}} = g Q_{\mib{i}}\rho_{\mib{i}},
\end{equation}
and
\begin{equation}
  \label{h2}
  H^{(2)}_{\mib{i}} = P_{\mib{i}}^2/(2M)+V(Q_{\mib{i}}),
\end{equation}
where $g$ is electron-vibration coupling constant,
$\rho_{\mib{i}}$ denotes local charge density given by
$\rho_{\mib{i}}=\sum_{\sigma}c^{\dagger}_{\mib{i}\sigma}c_{\mib{i}\sigma}$,
$c_{\mib{i}\sigma}$ is an annihilation operator of electron at site $\mib{i}$,
$Q_{\mib{i}}$ is normal coordinate of the oscillator,
$P_{\mib{i}}$ indicates the corresponding canonical momentum,
$M$ is mass of the oscillator,
and $V$ denotes an anharmonic potential for the oscillator, given by
\begin{equation}
  V(Q_{\mib{i}}) = k Q_{\mib{i}}^{2}/2+k_{4}Q_{\mib{i}}^{4}
                 + k_{6} Q_{\mib{i}}^6.
\end{equation}
Here $k$ denotes a spring constant, while $k_{4}$ and $k_{6}$ are
the coefficients for fourth- and sixth-order anharmonic terms, respectively.

Let us provide a comment on the present potential $V(Q_{\mib{i}})$
composed of second-, fourth-, and sixth-order terms.
It may be possible to prepare a simpler anharmonic potential
with negative second-order coefficient and positive $k_4$,
but we intend to use the potential with positive second-order coefficient,
since it seems to be natural to consider that the spring constant is taken
to be positive in the oscillation problem.
Then, in order to prepare the symmetric potential
which has a wide and flat region in the bottom,
we set negative $k_4$ and positive $k_6$ in the model potential.
For the case of $k_4=k_6=0$,
we immediately obtain the harmonic potential.
We believe that the present model is useful for the purpose to
grasp easily the effect of anharmonicity
in comparison with the results of the harmonic potential.
Note, however, that it is necessary to pay our attention
to the artificial aspects of the present model potential.

Now we define the phonon annihilation operator $a_{\mib{i}}$
at site $\mib{i}$ through the relation of
$Q_{\mib{i}} = (a_{\mib{i}}^{\dagger}+a_{\mib{i}})/\sqrt{2M\omega}$,
where $\omega$ denotes the energy of oscillation,
given by $\omega = \sqrt{k/M}$.
Then, we rewrite eqs.~(\ref{h1}) and (\ref{h2}), respectively, as
\begin{equation}
  H^{(1)}_{{\mib i}} = \sqrt{\alpha} \omega 
  (a_{\mib{i}}+a_{\mib{i}}^{\dagger}) \rho_{\mib{i}},
\end{equation}
and
\begin{equation}
  H^{(2)}_{{\mib i}} = \omega [a_{\mib{i}}^{\dagger}a_{\mib{i}}+1/2
  + \beta(a_{\mib{i}}+a_{\mib{i}}^{\dagger})^4
  + \gamma(a_{\mib{i}}+a_{\mib{i}}^{\dagger})^{6}],
\end{equation}
where $\alpha$ is non-dimensional electron-phonon coupling constant,
defined by
\begin{equation}
  \alpha = g^{2}/(2 M \omega^{3}),
\end{equation}
and $\beta$ and $\gamma$ are non-dimensional fourth- and sixth-order
anharmonicity parameters, given by
\begin{equation}
  \beta = k_{4}/(4 M^{2}\omega^{3}),~
  \gamma = k_{6}/(8M^{3}\omega^{4}).
\end{equation}
With the use of non-dimensional parameters $\alpha$, $\beta$, and $\gamma$,
it is convenient to rewrite the potential $V$ as
\begin{equation}
  V(q_{\mib{i}}) = \alpha \omega (q_{\mib{i}}^{2}
  + 16 \alpha \beta q_{\mib{i}}^{4}
  + 64 \alpha^{2} \gamma q_{\mib{i}}^{6}),
\end{equation}
where $q_{\mib{i}}$ is non-dimensional displacement defined by
$q_{\mib{i}}=Q_{\mib{i}}/\ell$ and the length scale $\ell$ is given by
$\ell = \sqrt{2\alpha/(M\omega)}$.

\subsection{Dependence of parameters on oscillator mass}

In order to discuss the isotope effect on $T_{\rm c}$,
let us define the $M$ dependence of parameters.\cite{Hotta1}
It is well known that $\omega$ is in proportion to $M^{-1/2}$
from $\omega = \sqrt{k/M}$,
when we assume that the spring constant $k$ does not depend on $M$.
If we further assume that $g$ is independent of $M$,
we obtain $\alpha \propto M^{1/2}$.
Concerning anharmonicity parameters $\beta$ and $\gamma$,
we obtain the relations of
$\beta \propto M^{-1/2}$ and $\gamma \propto M^{-1}$
by further assuming that $k_{4}$ and $k_{6}$ are independent of $M$.
We define $m$ as the mass ratio of the guest atom
due to the replacement by the isotope.
Then, we obtain $m$ dependence of parameters as
\begin{equation}
 \label{m-dep}
  \omega=\frac{\omega_0}{\sqrt{m}},~
  \alpha = \alpha_0 \sqrt{m},~
  \beta= \frac{\beta_0}{\sqrt{m}},~
  \gamma = \frac{\gamma_0}{m},
\end{equation}
where the subscript ``0'' denotes the quantity before we consider
the change of the oscillator mass.
Note that the length scale $\ell$ does not depend on $m$.

For the discussion of $T_{\rm c}$,
we define an electron-phonon coupling constant $\lambda$.
In general, we cannot obtain $\lambda$ analytically for anharmonic phonons,
but for the case of harmonic phonons ($\beta_0=\gamma_0=0$),
we simply obtain $\lambda=2\alpha \omega/W$
with electron bandwidth $W$.
As easily understood from the $m$ dependence of parameters,
$\lambda$ for harmonic phonons does not depend on $M$.
Thus, as a quantity to indicate the strength of electron-phonon coupling
for superconductivity, it is useful to define $\lambda_0$ as
\begin{equation}
  \lambda_0=2\alpha_0 \omega_0/W.
\end{equation}

Here we discuss the shape of the anharmonic potentials
considered in this paper.
As already mentioned in our previous papers,\cite{Oshiba-Proc,Hotta2}
the potential shapes are classified into three types:
On-center type for $-\sqrt{3\gamma /4} < \beta$,
rattling type for $-\sqrt{\gamma} < \beta < -\sqrt{3\gamma /4}$,
and off-center type $\beta < -\sqrt{\gamma}$.
Note that the range of $\beta$ to determine the potential type
depends on the value of $\gamma$.

In order to characterize the potential shape in the same parameter range,
we introduce the re-scaled anharmonicity parameter $\beta'$ as
$\beta'=\beta/\gamma$.
By using eqs.~(\ref{m-dep}), we easily obtain
\begin{equation}
  \beta'= \beta_0' = \beta_0/\sqrt{\gamma_0},
\end{equation}
indicating that $\beta'$ does not depend on $m$.
With the use of $\beta_{0}'$,
the potential shapes are classified into three types:
On-center type for $\beta'_0 > -\sqrt{3}/2$,
rattling type for $-1 < \beta'_0 < -\sqrt{3}/2$,
and off-center type $\beta'_0 < -1$.

\begin{figure}[t]
\includegraphics[width = 8.5cm]{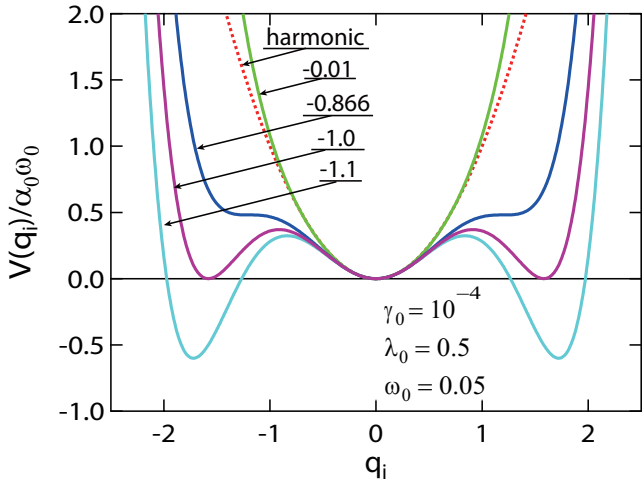}
\caption{(Color online) Anharmonic potentials for $\beta'_{0} = -0.1$,
$-0.866$, $-1.0$, and $-1.1$ for $\gamma_0 = 10^{-4}$, $\lambda_0 = 0.5$,
and $\omega_0 = 0.05$.
Dotted curve denotes harmonic potential.}
\end{figure}

In Fig.~1, we show the anharmonic potentials for several values of
$\beta'_0$ with $\gamma_0=10^{-4}$, $\omega_0=0.05$, and $\lambda_0=0.5$.
We note that the potential shapes are independent of $m$,
since we assume that $k$, $k_{4}$, and $k_{6}$ do not depend on $m$.
Throughout this paper, we set $\omega_{0} = 0.05$
in order to keep the adiabatic condition.
As for electron-phonon coupling constant $\lambda_0$,
we fix it as $\lambda_0=0.5$.

\subsection{Migdal-Eliashberg formalism}

In Ref.~\citen{Oshiba}, we have developed the strong-coupling theory
for superconductivity in the anharmonic Holstein model
by applying the formalism of the Migdal-Eliashberg theory
within the adiabatic region of $\omega \ll W$.
In order to make this paper self-contained, in this subsection,
we briefly explain the framework to evaluate
$T_{\rm c}$ in the strong-coupling region.

In the second-order perturbation theory in terms of $g$,
the linearized gap equation at $T=T_{\rm c}$ is given by
\begin{equation}
  \label{a-self}
  \phi(i \omega_{n}) = \alpha \omega^{2} T \sum_{n', {\mib k'}}
  D_{0}(i\omega_{n}-i\omega_{n'}) F({\mib k'},i\omega_{n'} ),
\end{equation}
where $\phi(i\omega_{n})$ is anomalous self-energy,
$\omega_{n}$ is fermion Matsubara frequency defined by
$\omega_{n} = (2n+1)\pi T$ with a temperature $T$ and an integer of $n$,
$D_{0}$ is bare phonon Green's function,
and $F$ is anomalous Green's function.
In the vicinity of $T_{\rm c}$, $F({\mib k}, i\omega_{n})$
is given in the linearized form as
\begin{equation}
  \label{a-green}
  F({\mib k}, i\omega_{n}) = -G({\mib k}, i\omega_{n})
  G(-{\mib k},-i\omega_{n})\phi(i\omega_{n}),
\end{equation}
where $G({\mib k}, i\omega_{n})$ is normal Green's function,
given by 
\begin{equation}
  \label{n-green}
  G({\mib k}, i\omega_{n}) = \frac{1}{i\omega_{n} - \varepsilon_{\mib k}
  -\Sigma(i\omega_{n})}.
\end{equation}
Here $\Sigma(i\omega_{n})$ is normal electron self-energy.
In the second-order perturbation theory in terms of $g$,
$\Sigma$ is expressed as
\begin{equation}
  \label{n-self}
  \Sigma(i\omega_{n}) = -\alpha \omega^{2} T \sum_{n', {\mib k}'}
  D_{0}(i\omega_{n}-i\omega_{n'}) G({\mib k}', i\omega_{n'}).
\end{equation}  
Since we consider Einstein-type local phonon, the site dependence
of $D_{0}$ does not appear and in the adiabatic approximation,
electron self-energy does not depend on momentum.

Concerning the phonon Green's function of the anharmonic phonon system,
we use $D_0$ instead of dressed phonon Green's function
by ignoring the phonon self-energy effect.
In the spectral representation, $D_{0}$ is given by
\begin{equation}
  D_{0}(i\nu_{n}) = \int \frac{\rho_{\rm ph}(\Omega)}{i\nu_{n}-\Omega} d\Omega,
\label{phonon_green}
\end{equation}
where $\nu_{n}$ is the boson Matsubara frequency defined
by $\nu_{n} = 2n\pi T$, and $\rho_{\rm ph}(\Omega)$ is
phonon spectral function, given by
\begin{equation}
  \rho_{\rm ph}(\Omega) = \sum_{K,L} A_{K,L}
  \delta(\Omega+E_K-E_L).
\end{equation}
Here $E_K$ is the $K$-th eigenenergy of $H^{(2)}_{\mib{i}}$
and the spectral weight $A_{K,L}$ is given by
\begin{equation}
 A_{K,L}= \frac{1}{Z}
  (e^{- E_{K}/T}-e^{- E_{L}/T})
  | \langle K|(a_{\mib i} + a_{\mib i}^{\dagger})|L \rangle |^{2},
\end{equation}
where $|K\rangle$ is the $K$-th eigenstate of $H^{(2)}_{\mib{i}}$ and
$Z$ is the partition function, given by $Z=\sum_K e^{-E_{K}/T}$.

In order to obtain $T_{\rm c}$, first we calculate the normal self-energy
$\Sigma$ by solving eqs.~(\ref{n-green}) and (\ref{n-self})
in a self-consistent manner.
Next we solve the gap equation eqs.~(\ref{a-self}) and (\ref{a-green})
by using $G$ in eq.~(\ref{n-green}).
Then, we obtain $T_{\rm c}$ as a temperature at which
the positive maximum eigenvalue of eq.~(\ref{a-self}) becomes unity. 
In actual calculations, we assume the electron density of states $1/W$
with rectangular shape of the electron bandwidth $W$.
Note here that $W$ is taken as the energy unit $W = 1$ in this paper.
For the sum on the imaginary axis, we use 32768 Matsubara frequencies.
Note also that we safely calculate $T_{\rm c}$ larger than $0.001$
for this number of Matsubara frequencies.
In order to accelerate the sum of large amount of Matsubara frequencies
in eqs.~(\ref{a-self}) and (\ref{n-self}),
we exploit the fast-Fourier-transformation algorithm.
For the evaluation of the eigenvalue of the gap equation eq.~(\ref{a-self}),
we use the power method.
Note again that we set $\omega_0 = 0.05$ and
$\lambda_0 = 2 \alpha_0 \omega_0 /W = 0.5$ in the calculations.

\section{Calculated Results}

\subsection{Superconducting transition temperature}

\begin{figure}[t]
\includegraphics[width = 8.5cm]{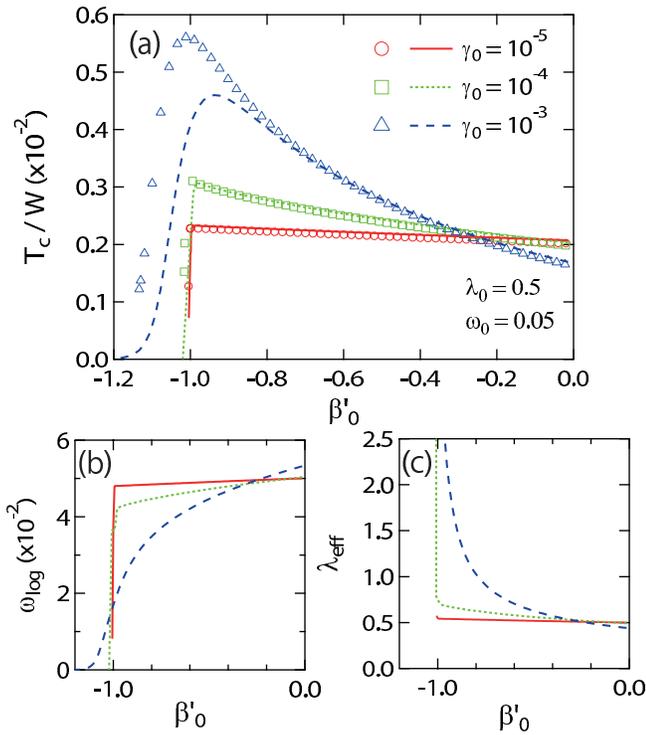}
\caption{(Color online)
(a) Superconducting transition temperature $T_{\rm c}$ vs. $\beta'_{0}$
for $\gamma_0$=$10^{-5}$, $10^{-4}$, and $10^{-3}$.
Curves indicate $T_{\rm c}$ calculated by the McMillan formula.
(b) Characteristic phonon energy $\omega_{\rm log}$ vs. $\beta'_{0}$.
(c) Effective electron-phonon coupling constant
$\lambda_{\rm eff}$ vs. $\beta_{0}'$}
\end{figure}

Before proceeding to the discussion on the isotope effect,
we exhibit the results on the superconducting transition temperature
$T_{\rm c}$ for the case of $m=1$.
In Fig.~2(a), we depict $T_{\rm c}$ vs. $\beta_{0}'$ by open symbols
for $\gamma_0=10^{-5}$, $10^{-4}$, and $10^{-3}$.
Note that sixth-order anharmonicity is the largest in the case of
$\gamma_{0} = 10^{-3}$, since it increases with the increase of $\gamma_{0}$.
Among the three curves for $T_{\rm c}$,
for $\gamma_{0}=10^{-4}$ and $10^{-3}$,
it is found that $T_{\rm c}$ increases with the decrease of $\beta_{0}'$
in the range of $\beta_{0}' > -1$
and it turns to be decreased at $\beta_{0}' = -1$.
Namely, the curve for $T_{\rm c}$ forms a peak structure around at
$\beta_0'= -1$.
The maximum value of $T_{\rm c}$ depends on $\gamma_0$.
In our previous work, the highest $T_{\rm c}$ has been found
for $\gamma_0 = 10^{-3}$.\cite{Oshiba}
For the case of $\gamma_{0}=10^{-5}$, we also observe that $T_{\rm c}$
increases with the decrease of $\beta_{0}'$ in the range of $\beta_{0}' > -1$,
but the rate of the increase is very slow.

Note that for small $\gamma_0$, it is found that $T_{\rm c}$ suddenly
decreases at $\beta_0'=-1.0$, because $T_{\rm c}$ is strongly influenced
by the change of the anharmonic potential
from rattling- to off-center type, as observed in Fig.~1.
The calculation of $T_{\rm c}$ in the present approximation is not considered
to be valid for the off-center type potential, since double degeneracy
in the phonon energy affects seriously on the low-energy electron states.
This point will be discussed later again in Sec.~4, but in any case,
the extension of the calculation to the off-center type potential is
one of our future problems.

In order to understand the formation of the peak in $T_{\rm c}$,
we consider the McMillan formula.\cite{McMillan}
The McMillan formula of $T_{\rm c}$ which we should analyze is given by
\begin{equation}
 T_{\rm c}^{\rm Mc} = \frac{\omega_{\log}}{1.20} {\rm exp}
 \left( -\frac{1+\lambda_{\rm eff}}{\lambda_{\rm eff}} \right),
\label{McMillan}
\end{equation}
where the effective Coulomb interaction is simply ignored
in the present calculations,
but this point will be also discussed later in Sec.~4.
Note that we replace a numerical factor $1.04$ in front of
$1+\lambda_{\rm eff}$ in the original formula with the unity
by following Allen and Dyne.\cite{Allen}
Here $\lambda_{\rm eff}$ indicates
the effective electron-phonon coupling constant, given by
\begin{eqnarray}
   \label{lambdaeff}
   \lambda_{\rm eff} = \frac{\lambda \omega}{2} \int^{\infty}_{-\infty}
   \frac{\rho_{\rm ph}(\Omega)}{\Omega} \, d\Omega,
\end{eqnarray}
and $\omega_{\log}$ indicates the characteristic phonon energy
defined by
$\omega_{\rm log} = {\rm exp}( \langle \log \Omega \rangle)$,
where $\langle \log \Omega \rangle$ is given by
\begin{eqnarray}
  \label{omegalog}
  \langle \log \Omega \rangle = \frac{\lambda \omega}{\lambda_{\rm eff}}
  \int_{0}^{\infty}\frac{\rho_{\rm ph}(\Omega)}{\Omega}{\log}\, \Omega \,d\Omega.
\end{eqnarray}
Note that $\omega_{\log}$ and $\lambda_{\rm eff}$ depend on $T$,
since $\rho_{\rm ph}(\Omega)$ includes the Boltzmann factor.
Namely, eq.~(\ref{McMillan}) becomes the self-consistent equation
for $T_{\rm c}$.
Thus, we define $T_{\rm c}^{\rm Mc}$ as a temperature at which
the left- and right-hand terms of eq.~(\ref{McMillan})
are equal to each other.

In Fig.~2(a), we depict $T_{\rm c}^{\rm Mc}$ as functions of $\beta_{0}'$
for $\gamma_{0} = 10^{-5}$, $10^{-4}$, and $10^{-3}$.
For small $\gamma_0$ such as $\gamma_{0} = 10^{-5}$,
it is difficult to perform the self-consistent calculation of $T_{\rm c}^{\rm Mc}$
in the vicinity of $\beta_0'=-1.0$, when $T_{\rm c}^{\rm Mc}$ becomes very low.
However, for $\gamma_{0} = 10^{-5}$ and $10^{-4}$,
it is found that in the wide range of $\beta_0'$,
$T_{\rm c}^{\rm Mc}$ well reproduces $T_{\rm c}$
obtained by the Eliashberg equation.
For $\gamma_0 = 10^{-3}$, $T_{\rm c}^{\rm Mc}$ is similar to the solution
of the Eliashberg equation in the region of $\beta_{0}' > -0.8$,
while in $\beta_{0}' < -0.8$, the magnitude of $T_{\rm c}^{\rm Mc}$
is different from that of $T_{\rm c}$,
although $T_{\rm c}^{\rm Mc}$ qualitatively exhibits the behavior
with the peak formation in $T_{\rm c}$.

In Figs.~2(b) and 2(c), we show $\omega_{\rm log}$ and $\lambda_{\rm eff}$
at $T = T_{\rm c}^{\rm Mc}$.
For harmonic phonons,
we obtain $\omega_{\rm log}=\omega$ and $\lambda_{\rm eff}=\lambda_{0}$
from the spectral function of
$\rho_{\rm ph}(\Omega) = \pm \delta (\Omega \mp \omega)$.
For $\gamma_0 = 10^{-5}$ and $10^{-4}$,
$\omega_{\rm log}$ is almost equal to $\omega(=0.05)$
in the range of $\beta_{0}' > -0.2$.
We also observe $\lambda_{\rm eff} \approx \lambda_{0} (=0.5)$
in the same range of $\beta_{0}'$.
It is considered that the guest ion exhibits the harmonic oscillation
around at the origin of the potential.
Since $\omega_{\rm log}$ and $\lambda_{\rm eff}$ moderately deviate
from $\omega$ and $\lambda$, respectively, for $-1.0 < \beta_{0}' < -0.2$,
anharmonicity slightly affects them and
$T_{\rm c}$ slowly increases with the decrease of $\beta_{0}'$.
For $\beta_{0}' < -1.0$, since the decrease of $\omega_{\rm log}$
and the increase of $\lambda_{\rm eff}$ are very rapid,
it is considered that $T_{\rm c}$ rapidly decreases.
Note that for $\gamma_{0} = 10^{-5}$,
it is difficult to obtain reliable solutions of eq.~(\ref{McMillan})
with eqs.~(\ref{lambdaeff}) and (\ref{omegalog})
at $\beta_{0}' \simeq -1.0$,
since $\lambda_{\rm eff}$ and $\omega_{\rm log}$ depend so sensitively
on $\beta_0'$ at the region.

For $\gamma_{0} = 10^{-3}$, $\omega_{\rm log}$ and $\lambda_{\rm eff}$ are
significantly different from harmonic results
for the wide region of $\beta_{0}'$,
since the dependence of eigenenergies on $\beta_{0}'$ is rather different
from that of harmonic phonons for large $\gamma$.\cite{Oshiba}
We observe that the value of $T_{\rm c}$ changes in the wide region of
$\beta_{0}'$ for $\gamma_{0}= 10^{-3}$.
With the decrease of $\beta_{0}'$, $\omega_{\log}$ decreases and
$\lambda_{\rm eff}$ increases monotonically.
It is understood that the peak of $T_{\rm c}$ is formed due to
the competition of decreasing $\omega_{\rm log}$ and increasing
$\lambda_{\rm eff}$.

\begin{figure}[t]
\includegraphics[width = 8.5cm]{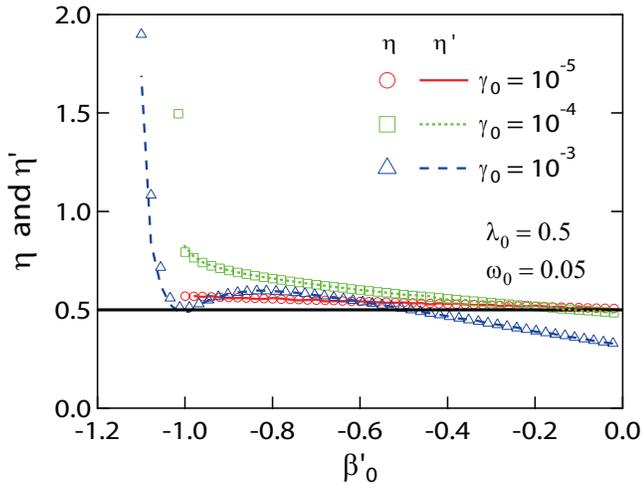}
\caption{(Color online) Exponent $\eta$ vs. $\beta'_{0}$
(open symbols) obtained from the solution of the Eliashberg equation
for $\gamma_{0} =  10^{-5}$, $10^{-4}$, and $10^{-3}$.
Curves indicate the exponent given by
$\eta' = \eta_{\alpha} + \eta_{\omega} + \eta_{\beta} + \eta_{\gamma}$.
The horizontal line indicates $\eta = 1/2$ of the normal isotope effect.}
\end{figure}

\subsection{Exponent of the isotope effect}

Now we consider the exponent $\eta$ of the isotope effect,
which is evaluated by
\begin{equation}
  \label{eta}
  \eta=-\frac{d \log T_{\rm c}}{d \log m}.
\end{equation}
Since we cannot analytically calculate $\eta$,
the derivative in eq.~(\ref{eta}) is approximated by
the differentiation in terms of $m$.
Namely, $d \log T_{\rm c}/d \log m$ is numerically estimated as
\begin{equation}
 \frac{d \log T_{\rm c}}{d \log m} =
 \frac{\log T_{\rm c}(m= 1.01) - \log T_{\rm c}(m = 1)}{\log 1.01},
\end{equation}
where $T_{\rm c}(m)$ denotes $T_{\rm c}$ for the case of mass ratio $m$.

In Fig.~3, we depict $\eta$ vs. $\beta_{0}'$ by open symbols
for $\gamma_{0}$=$10^{-5}$, $10^{-4}$, and $10^{-3}$.
For $\gamma_{0}=10^{-5}$ and $10^{-4}$,
at $\beta_{0}'=0$, $\eta$ is almost equal to $1/2$ which is
the value for harmonic phonons.
These results seem to be natural, when we recall
that the effect of anharmonicity is weak at $\beta_{0}'=0$,
as observed in Fig.~1 for the potential shape.
When $\beta_{0}'$ is decreased, $\eta$ slowly increases
in the range of $\beta_{0}' > -1$ and
it rapidly increases in the region of the off-center type potential.
For $\gamma_0=10^{-3}$, $\eta$ is less than $1/2$
in the range of $0 > \beta_{0}' > -0.5$.
However, $\eta$ is larger than $1/2$ for $\beta_{0}' < -0.5$ and
it also rapidly increases in region of the off-center type potential
through the broad peak around at $\beta_0' \simeq -0.8$.

In order to clarify the origin of $\eta$ larger than $1/2$,
we decompose $\eta$ into four parts as
\begin{eqnarray}
\label{etac}
 \eta'=\eta_{\alpha}+\eta_{\omega}+\eta_{\beta}+\eta_{\gamma},
\end{eqnarray}
where $\eta_{\alpha}$, $\eta_{\omega}$, $\eta_{\beta}$, and $\eta_{\gamma}$
are given by
\begin{equation}
\begin{split}
\label{etas}
\eta_{\alpha} &=  -\frac{\partial \log \alpha}{\partial \log m}\frac{\partial
 \log T_{\rm c}}{\partial \log \alpha} =  -\frac{1}{2}\frac{\partial
 \log T_{\rm c}}{\partial \log \alpha},\\
\eta_{\omega} &=  -\frac{\partial \log \omega}{\partial \log m}\frac{\partial
 \log T_{\rm c}}{\partial \log \omega} = \frac{1}{2}\frac{\partial
 \log T_{\rm c}}{\partial \log \omega},\\
\eta_{\beta} &=  -\frac{\partial \log \beta}{\partial \log m}\frac{\partial
 \log T_{\rm c}}{\partial \log \beta} = \frac{1}{2}\frac{\partial
 \log T_{\rm c}}{\partial \log \beta}, \\
\eta_{\gamma} &=  -\frac{\partial \log \gamma}{\partial \log m}\frac{\partial
 \log T_{\rm c}}{\partial \log \gamma} = \frac{\partial
 \log T_{\rm c}}{\partial \log \gamma},
\end{split}
\end{equation}
respectively.
Note that we consider $\alpha$, $\omega$, $\beta$, and $\gamma$
as variables of $T_{\rm c}$ and $\eta$.
In order to distinguish $\eta$'s of eq.~(\ref{eta}) and eq.~(\ref{etac}),
we use the notation of $\eta'$ in eq.~(\ref{etac}).

It is instructive to evaluate eq.~(\ref{etac})
for the case of harmonic phonons.
By calculating eqs.~(\ref{etas}) with the use of eq.~(\ref{McMillan})
for the harmonic case, we obtain
\begin{equation}
 \eta_{\alpha} = 1/2-\eta_{\omega} = -1/(2\lambda_0),~
 \eta_{\beta} =\eta_{\gamma}=0.
\end{equation}
Thus, irrespective of the value of $\lambda_0$,
we obtain $\eta_{\alpha}+\eta_{\omega}=1/2$,
which is just the exponent of the normal isotope effect.

As for the evaluation of eqs.~(\ref{etas}) for anharmonic phonons,
each derivative is also approximated by the differentiation.
Then, we numerically estimate $\partial \log T_{\rm c}/ \partial \log x$ as
\begin{eqnarray}
 \frac{\partial \log T_{\rm c}}{\partial \log x}
 =\frac{\log T_{\rm c}(x+\Delta x)-\log T_{\rm c}(x)}{\log (x+\Delta x)-\log x},
\end{eqnarray}
where $T_{\rm c}(x)$ indicates $T_{\rm c}$ as a function of $x$,
$\Delta x$ indicates small deviation of $x$, and
$x$ denotes the variable among $\alpha$, $\omega$, $\beta$, and $\gamma$.
In order to obtain enough precision of the numerical differentiation,
we choose $\Delta x/x=0.1\% \sim 0.001\%$.
At $\beta_{0}' \simeq -1$, we set $\Delta x/x = 0.001\%$ for $\gamma_{0} = 10^{-4}$,
but it is difficult to obtain reliable values for $\beta_{0}'<-1$.
For $\gamma_{0} = 10^{-5}$, it is also difficult to evaluate $\eta$
with enough precision for $\beta_{0}' \leq -1$,
even if we set $\Delta x/x= 0.001\%$.
In Fig.~3, we depict $\eta'$ by curves on the open symbols of $\eta$
for $\gamma_{0} = 10^{-5}$, $10^{-4}$, and $10^{-3}$.
Note that for $\gamma_{0} = 10^{-5}$ and $10^{-4}$,
we do not show the curves in the region of $\beta_0' \leq -1.0$,
since we could not evaluate each term of eqs.~(\ref{etas})
with enough precision.
At $\beta_{0}' \simeq -1$, since $\eta_{\beta}$ and $\eta_{\gamma}$ are
very sensitive for anharmonicity, evaluation is difficult.
However, it is considered that
$\eta'$ agrees well with $\eta$ within the numerical error-bars.

\begin{figure}[t]
\includegraphics[width = 8.5cm]{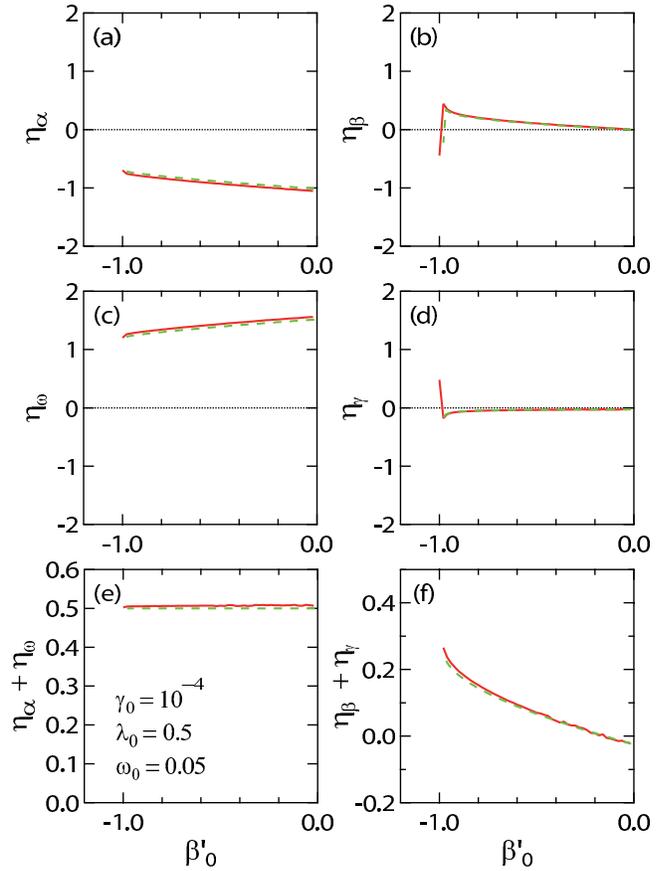}
\caption{(Color online) The parts of $\eta'$ vs. $\beta_{0}'$
for $\gamma = 10^{-4}$.
(a) $\eta_{\alpha}$ vs. $\beta_{0}'$.
(b) $\eta_{\beta}$ vs. $\beta_{0}'$.
(c) $\eta_{\omega}$ vs. $\beta_{0}'$.
(d) $\eta_{\gamma}$ vs. $\beta_{0}'$.
(e) $\eta_{\alpha} + \eta_{\omega}$ vs. $\beta_{0}'$.
(f) $\eta_{\beta} + \eta_{\gamma}$ vs. $\beta_{0}'$.
In each panel, solid and broken curves indicate the results of
the Eliashberg equation and the McMillan formula, respectively.}
\end{figure}

In Figs.~4(a)-4(d), we show $\eta_{\alpha}$, $\eta_{\beta}$, $\eta_{\omega}$,
and $\eta_{\gamma}$, respectively, as functions of $\beta_{0}'$.
In Figs.~4(e) and 4(f), we depict $\eta_{\alpha}+\eta_{\omega}$ and
$\eta_{\beta}+\eta_{\gamma}$, respectively.
In addition, we also show the parts of $\eta'$ evaluated from
the McMillan formula of $T_{\rm c}$ by broken curves in Figs.~4(a)-4(f).
We observe that each term of the exponent $\eta'$ evaluated from
the McMillan formula reproduces well the corresponding result of
the Eliashberg equation.
Thus, it is possible to discuss the behavior of $\eta$ and $\eta'$
on the basis of the McMillan formula.

In Figs.~4(a) and 4(c), $\eta_{\alpha}$ and $\eta_{\omega}$ are shown.
At $\beta_{0} = 0$, we find
$\eta_{\alpha} \approx -1$ and $\eta_{\omega} \approx 1.5$,
which are the values for the harmonic phonons with $\lambda_0=0.5$.
With the decrease of $\beta_{0}'$, $\eta_{\alpha}$ increases and
$\eta_{\omega}$ decreases due to the effect of anharmonicity,
but the sum of $\eta_{\alpha} + \eta_{\omega}$ is still almost $0.5$
for the wide range of $\beta_{0}'$, as observed in Fig.~4(e).
Namely, as long as the anharmonicity is not so strong,
the exponent $1/2$ of the normal isotope effect originates from
$\eta_{\alpha} + \eta_{\omega}$ even for the anharmonic potentials.

In Figs.~4(b) and 4(d), we depict $\eta_{\beta}$ and $\eta_{\gamma}$.
They are almost zero at $\beta_{0}' = 0$, when anharmonicity is weak.
With the decrease of $\beta_{0}'$, $\eta_{\beta}$ increases and
$\eta_{\gamma}$ decreases.
Note that $\beta$ plays a role to expand the width of anharmonic
potential, as observed in Fig.~1.
Thus, $\eta$ is enhanced with the increase of the amplitude of the
guest ion for $\beta_{0}' > -1.0$.
For $\beta_{0} < -1.0$, since the potential shape is suddenly changed to the
off-center type, $\eta_{\beta}$ is also suddenly changed,
but we are not interested in such behavior at the present stage.
On the other hand, we note that $\gamma$ plays a role to reduce
the width of anharmonic potential.
The effect of $\gamma$ decreases the amplitude of the guest ion
and it also decreases $\eta$ for $\beta_{0}' > -1.0$.

In Fig.~4(f), we depict $\eta_{\beta} + \eta_{\gamma}$,
which increases with the decrease of $\beta_{0}'$.
At $\beta_{0}' \sim 0$, since the fourth- and sixth-order anharmonicity
are very small, $\eta_{\beta} + \eta_{\gamma}$ is almost zero.
For $\gamma_{0} = 10^{-4}$,
the sixth-order anharmonicity is moderately strong
and $\eta_{\beta} + \eta_{\gamma}$ slightly deviates from zero.
The behavior of $\eta_{\beta} + \eta_{\gamma}$ is quite similar to
that of $\eta-1/2$.
In short, it is found that $\eta_{\alpha} + \eta_{\omega} \approx 1/2$ and
$\eta_{\beta} + \eta_{\gamma}$ determines the deviation of $\eta$ from $1/2$.
We consider that $\eta_{\alpha} + \eta_{\omega}$ represents the normal
isotope effect of $\eta=1/2$,
while $\eta_{\beta} + \eta_{\gamma}$ indicates the effect of anharmonicity
on the exponent of the isotope effect.

\begin{figure}[t]
\includegraphics[width = 8.5cm]{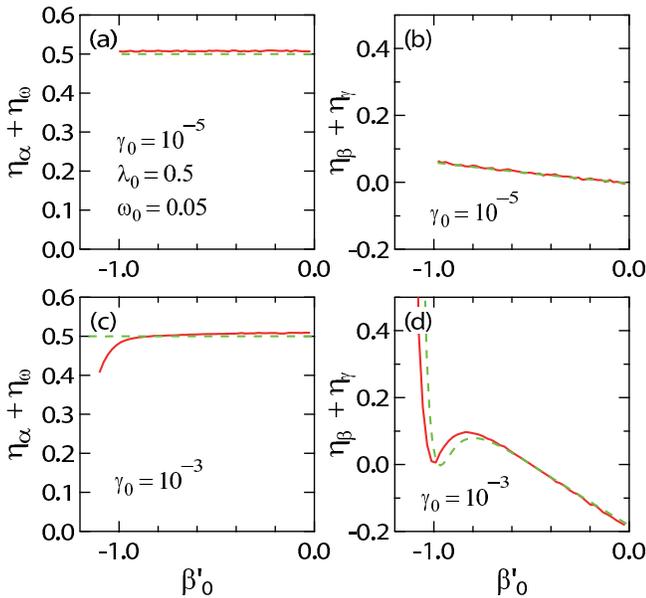}
\caption{(Color online)
(a) $\eta_{\alpha} + \eta_{\omega}$ vs. $\beta_{0}'$ for $\gamma=10^{-5}$.
(b) $\eta_{\beta} + \eta_{\gamma}$ vs. $\beta_{0}'$ for $\gamma=10^{-5}$.
(c) $\eta_{\alpha} + \eta_{\omega}$ vs. $\beta_{0}'$ for $\gamma=10^{-3}$.
(d) $\eta_{\beta} + \eta_{\gamma}$ vs. $\beta_{0}'$ for $\gamma=10^{-3}$.
In each panel, solid and broken curves indicate the results of
the Eliashberg equation and the McMillan formula, respectively.}
\end{figure}

In Figs.~5(a)-5(d), we depict $\eta_{\alpha} + \eta_{\omega}$ and
$\eta_{\beta} + \eta_{\gamma}$ for $\gamma_{0} =10^{-5}$ and $10^{-3}$.
We again observe that $\eta_{\alpha} + \eta_{\omega} \approx 1/2$ and
the behavior of $\eta_{\beta} + \eta_{\gamma}$ determines the deviation
of $\eta$ from $1/2$,
except for the results in the vicinity of $\beta_0' =-1.0$ for $\gamma_0=10^{-3}$.
It is considered that $\eta > 1/2$ in the rattling-type potential
is mainly caused by the anharmonicity.

Here we provide a comment on non-monotonic behavior of $\eta$ in Fig.~4
for $\gamma_{0} = 10^{-3}$ and $-1.0 < \beta_{0}' <-0.8$.
Since $\eta_{\alpha} + \eta_{\omega}$ is almost equal to $1/2$
even for the case of $\gamma_{0} = 10^{-3}$, such non-monotonic behavior
originates from the anharmonicity part $\eta_{\beta} + \eta_{\gamma}$,
as observed in Fig.~5(d).
Roughly speaking, we consider that the peak structure is formed by the competition of
increasing $\eta_{\beta}$ and decreasing $\eta_{\gamma}$
when $\beta_0'$ is decreased.
Note that we do not further pursue the origin of each behavior of
$\eta_{\beta}$ and $\eta_{\gamma}$ at the present stage,
since it will depend on the anharmonic potential.
It is emphasized here that the behavior of $\eta_{\beta} + \eta_{\gamma}>0$
characterizes the anomalous exponent $\eta>1/2$ in the region of
rattling-type potential.

\section{Discussion and Summary}

In this paper, we have evaluated the exponent $\eta$ of the isotope effect
for rattling-induced superconductor in the strong-coupling analysis
by solving the gap equation of the Migdal-Eliashberg theory.
First, we have obtained that $\eta$ is larger than $1/2$ due to the
increase of anharmonicity in the region of the rattling-type potential.
Next, we have investigated the origin of $\eta > 1/2$ by evaluating four parts
of $\eta$ with the use of the chain rule of the derivative.
It has been clearly shown that the deviation of $\eta$ from $1/2$ is
due to the anharmonicity.
Then, we have considered that $\eta > 1/2$ can be the evidence of
rattling-induced superconductor.

Here we discuss the reliability of the result in the off-center type
potential with $\beta_{0}' < -1.0$.
For the purpose, we focus on the validity of the adiabatic approximation
in such a region.
We consider the adiabatic approximation as $\omega \ll W$,
but in the present calculation, we have found that
$\lambda_{\rm eff}$ monotonically increases
with the decrease of $\beta_{0}'$.\cite{Oshiba}
The increase of $\lambda_{\rm eff}$ indicates the strong-coupling
tendency, leading to the reduction of the effective bandwidth $W^*$.
Namely, $W^*$ decreases with the decrease of $\beta_{0}'$.
Here we note that $\lambda_{\rm eff}$ is rapidly enhanced
in the off-center type potential region.
Even if $W$ is much larger than the phonon energy $\omega$,
$W^*$ eventually becomes comparable with $\omega$,
leading to the violation of the adiabatic condition
in the off-center type potential region.
Thus, it is necessary to recognize that the results in the region
of the off-center type potential are not reliable
even in the strong-coupling analysis.
It is one of our future problems to develop a theory to consider
non-adiabatic effect through the electron-phonon vertex corrections
in the region of the off-center type potential.

Let us briefly discuss the effect of the Coulomb interaction,
which has been perfectly ignored in the present model.
In the famous McMillan formula,\cite{McMillan} $\eta$ is expressed by
\begin{eqnarray}
  \label{etaC}
  \eta = \frac{1}{2}\left[1-
  \frac{(1+\lambda)(1+0.62\lambda)\mu^{*2}}{\lambda +\mu^{*}(1+0.62\lambda)}
  \right],
\end{eqnarray}
where $\mu^{*}$ denotes the non-dimensional effective Coulomb interaction,
given by $\mu^*=(U/W)/[1+(U/W)\log (W/\omega_0)]$ with
the short-range Coulomb repulsion $U$.
From this expression, we easily understand that $\eta$ becomes smaller
than $1/2$, as observed in actual materials,
when we include the effect of the Coulomb interaction.
In this sense, our result of $\eta > 1/2$ is peculiar and it can be
the evidence for superconductivity induced by anharmonic phonons.

For $\beta$-pyrochlore oxides, electron-phonon coupling constant is larger
than about $0.8$ and $\mu^{*}$ is considered to be about $0.1.$\cite{Hiroi2}
If we simply use eq.~(\ref{etaC}) for the evaluation of $\eta$,
we obtain $\eta \approx 0.49$ and the reduction from $0.5$ is very small.
It is true that $\eta$ is reduced when we include the effect of the Coulomb
interaction, but in $\beta$-pyrochlore oxides,
we imagine that the effect of the Coulomb interaction is not strong
enough to reduce significantly the value of $\eta$.
In fact, recent de Haas-van Alphen oscillation measurements of
KOs$_2$O$_6$ have clearly suggest that the mass enhancement of
quasi-particle originates
from the electron-rattling interaction and the effect of
the Coulomb interaction is considered to be small.\cite{Terashima}

Finally, we provide a brief comment on the parameter region
corresponding to actual $\beta$-pyrochlore oxides.
We expect that the parameters of $\gamma_{0} = 10^{-3}$ and
$-1.0 < \beta_{0}' < -0.6$ correspond to $\beta$-pyrochlore oxides,
because the potential for those parameters exhibits
the flat and wide region at the bottom,
leading to rattling oscillation which will enhance $T_{\rm c}$.
However, there are insufficient evidences to prove
such correspondence at the present stage.
In order to discuss actual materials quantitatively
on the basis of our scenario,
it is necessary to develop further our studies in future.

In summary, we have found that the isotope effect
with the exponent $\eta > 1/2$ occurs for superconductivity
due to electron-rattling interaction.
From the detailed analysis of $\eta$, we have confirmed that the
deviation of $\eta$ from $1/2$ originates from the anharmonicity.
It is highly expected that the detect of this anomalous isotope effect
can be the evidence of superconductivity induced by rattling
in $\beta$-pyrochlore oxides.

\section*{Acknowledgement}

This work has been supported by a Grant-in-Aid for Scientific Research
on Innovative Areas ``Heavy Electrons''
(No. 20102008) of The Ministry of Education, Culture, Sports,
Science, and Technology, Japan.


\end{document}